\documentclass{elsart}
\usepackage{epsf}
\usepackage{epsfig}
\usepackage{cite}
\usepackage{url}

\usepackage[
 pdfauthor={S. Chekanov},
 pdftitle={RunMC},
 pdfsubject={an object-oriented framework for Monte Carlo models},
 pdfkeywords={Monte Carlo simulation, high-energy physics, C/C++}
 ]{hyperref}
\begin{document}

\bibliographystyle{elsart-num}

\begin{frontmatter}

\title{
RunMC - 
an object-oriented analysis framework for 
Monte Carlo simulation of high-energy particle collisions
}  

\author{S.~Chekanov}

% \maketitle

\address{HEP division, Argonne National Laboratory,
9700 S.Cass Avenue, \\
Argonne, IL 60439
USA \\
Email: chekanov@mail.desy.de}

% \begin{center}
% HEP division, Argonne National Laboratory,
% 9700 S.Cass Avenue, \\
% Argonne, IL 60439
% USA
% \end{center} 

\begin{abstract}
RunMC is an object-oriented framework aimed to
generate and to analyse high-energy collisions 
of elementary particles 
using Monte Carlo simulations. 
This package, being based on C++
adopted by CERN as the main programming 
language for the LHC experiments,  
provides a common interface to different
Monte Carlo models using modern physics
libraries. Physics calculations (projects) can easily be 
loaded and saved as external modules.
This simplifies the development of complicated calculations
for high-energy physics in large collaborations. 
This desktop program is open-source licensed and is available on
the LINUX and Windows/Cygwin platforms.
\end{abstract}


\end{frontmatter}

%%%%%%%%%%%%%%%%%%%%%%%%%%%%%%%%%%%%%%%%%%%%%%%%%%%%%%%%%%%%%%%%%
\section{Tabular Summary }
%%%%%%%%%%%%%%%%%%%%%%%%%%%%%%%%%%%%%%%%%%%%%%%%%%%%%%%%%%%%%%%%%%

\begin{tabular}{ll}
Program name: & RunMC  \\
Version:      & 3.3 \\
Date of last version: &  February, 2005 \\
Author: &  Sergei Chekanov \\
Size: &  15M (Linux/Windows-Cygwin), 12M (Windows-Cygwin) \\
Operating system: &  Linux, Windows/Cygwin \\
Additional packages needed: &  CLHEP, ROOT, CERNLIB with PDFLIB \\
Program requirements: &  g77, g++, make, X11, Java JRE1.4 and higher \\
Programming language: &  C++, C, Fortran, Java, bash \\
User manuals:  &  integrated into RunMC GUI \\
RMC files   :  &  integrated into RunMC GUI \\
Program availability:  &  \url{http://www.desy.de/~chekanov/runmc/} \\ 
                      &   \url{http://www.hep.anl.gov/chakanau/runmc/} 
\end{tabular}

%%%%%%%%%%%%%%%%%%%%%%%%%%%%%%%%%%%%%%%%%%%%%%%%%%%%%%%%%%%%%%%%%%
\section{Introduction}
%%%%%%%%%%%%%%%%%%%%%%%%%%%%%%%%%%%%%%%%%%%%%%%%%%%%%%%%%%%%%%%%%%

General-purpose Monte Carlo (MC) models for 
high-energy collisions 
of elementary particles are important tools 
used by theorists and experimentalists in their research.
At present, the MC models written in FORTRAN are 
widely used in many high-energy physics laboratories
worldwide (CERN, HERA, Fermilab etc.). 
These models are known to be fast, robust and well tested.

However, the main choice for future high-energy experiments 
is an object-oriented programming language, either 
C++ (the LHC experiments at CERN) or Java (the NLC project). 
Recently, a few steps towards converting the FORTRAN MC models 
to the C++ programming language 
were already undertaken \cite{Bertini:2000uh, ThePEG}. 
Such rewriting can take some time, since complicated physics
codes written in the old-fashion FORTRAN cannot be easily converted 
to the modern object-oriented programming language.
Furthermore, the MC models written in C++ require a thoughtful
verification to insure
that their predictions are consistent with the original FORTRAN-based
MC programs, as well as with the physics results obtained in the past.  
Such verifications will go over certain time, and a tool which allows 
to perform such comparisons is urgently needed.
  
Clearly, a program which allows 
running of  both FORTRAN-coded and
C++ MC models using a common
C++ programming environment should be valuable.
This is important not only 
for comparisons and verifications of these MC models. 
Such C++ framework can also extend the lifetime of 
FORTRAN-based models 
especially for the LHC, NLC and TEVATRON communities, and can  
provide  compatibility of most popular 
MC models with the new software to be used in the future.
In this approach, the MC output should be  
converted to C++ classes for further 
analysis or graphical representation (histograms).

The RunMC package provides this possibility. 
In addition, 
the graphical user interface (GUI) of this package was designed to be
as user 
friendly as possible. It helps 
to initialise the MC models and the user calculations, as well as to 
monitor the event generation. 
It also  provides a significant flexibility to
fill histograms and, at the same time, 
different MC models can be generated without changing the 
user code and histogram settings.
In future, Monte Carlo models based on C++ can also be included.
Thus, RunMC will provide a unified approach to 
generate and analysis 
very different MC models independent of their native codes.

The RunMC program
fully complies with the change
in the programming paradigm of data analysis.
Instead of the FORTRAN-based analysis tools, 
such as PAW \cite{paw} and HBOOK \cite{hbook}, 
it uses the modern 
CERN C++ analysis packages, 
CLHEP \cite{Lonnblad:1994kt,Fischler:2000em} 
and ROOT \cite{Brun:1997pa,Brun:2003ga}. 
Therefore, the program meets the requirements
of future high-energy experiments.

In this respect, the RunMC program is similar to the
JetWeb server \cite{JetWeb},  
which also provides the ability to compare
the existing MC models and to confirm the physics assumptions they contain.
However, in contrast to JetWeb, the RunMC program was designed as a standalone
desktop application. Therefore,
the user has full access to his calculations and to the program itself.

The RunMC package is not only the common C++ frond-end of 
Monte Carlo models designed to work with the modern analysis 
and graphic environment. Within this approach, the concept of 
project modules was introduced.
A project file, which can contain external 
calculations, MC tunings, histogram definitions, etc. 
can be loaded to RunMC with the same ease as
a document can be opened in the Microsoft Word 
program. 
Due to such simplicity, the program can be used for educational purposes,
since very different aspects of the hadronic final states can easily
be studied in a few mouse clicks.  
The project files are small and platform
independent, therefore, it is fairly simple to 
share complicated physics calculations between 
scientists in large collaborations.
 
%%%%%%%%%%%%%%%%%%%%%%%%%%%%%%%%%%%%%%%%%%
\section{The structure of RunMC}
%%%%%%%%%%%%%%%%%%%%%%%%%%%%%%%%%%%%%%%%%%

The RunMC package consists of the two parts: RunMC GUI 
and RunMC MC programs.
There are two implementations of RunMC GUI: one is written using
the Wide Studio C++ classes \cite{Hirabayashi}, and an alternative GUI 
based on Java.  

\begin{figure}
\includegraphics[height=10.0cm]{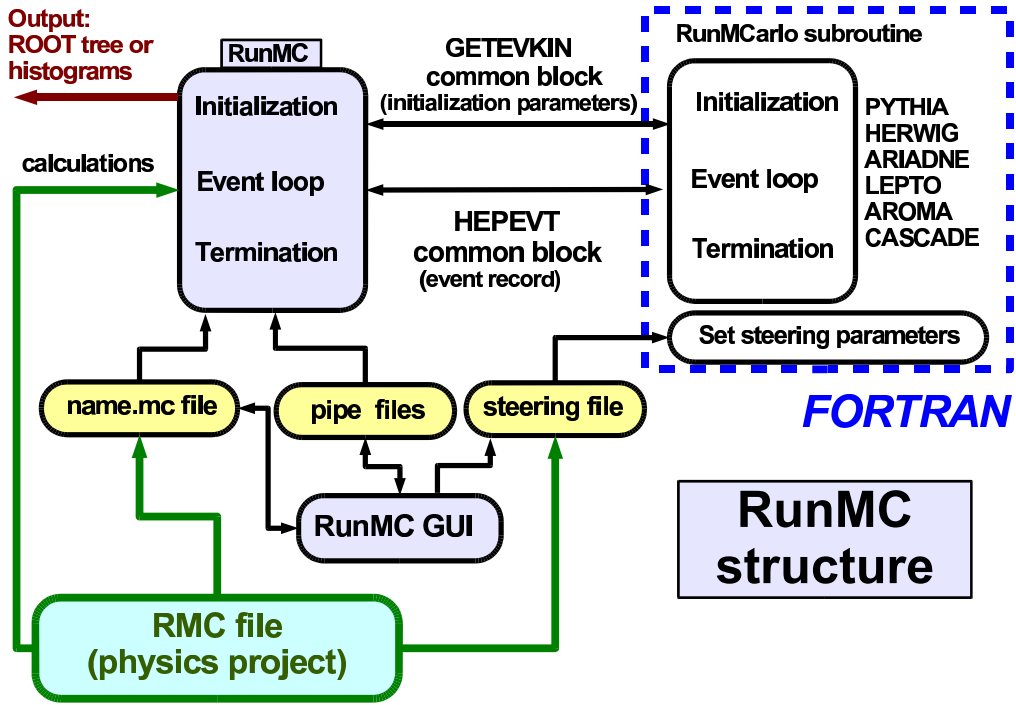}
\caption{
\label{structure}
The diagram shows the structure of the RunMC program.}
\end{figure}

The RunMC MC programs are 
based on the ROOT and CLHEP packages 
integrated with 
the FORTRAN MC models. 
A schematic structure of 
RunMC is illustrated in Fig.~\ref{structure}.

At present, the following MC models 
are included: PYTHIA~6.2~\cite{Sjostrand:2001yu}, 
HERWIG~6.5~\cite{herwig}, 
ARIADNE~4.12~\cite{Lonnblad:1992tz},
LEPTO~6.5~\cite{Ingelman:1996mq}, 
AROMA~2.2~\cite{Ingelman:1996mv}, 
CASCADE~1.2~\cite{Jung:2001hx},
PHOJET~1.05~\cite{phojet},    
RAPGAP~3.1~\cite{rapgap}.
Thus, there are several executable 
RunMC MC programs corresponding to each such MC model.

It should be pointed out that many hard processes included
in such models come from other authors. In future 
certain physics algorithms can be replaced, 
and new  processes can be added~\cite{genetic}.  
The RunMC package should reflect such a modularisation 
as much as possible. Therefore, an attempt will be made  
to accommodate this trend.

RunMC GUI communicates with the  RunMC MC programs using pipe files
located in a directory called ``\$RUNMC/pipes''. 
Here, ``\$RUNMC'' denotes the installation directory 
of the RunMC project.
All the directories to be discussed below are assumed to
be located in this directory.  

The RunMC MC programs and RunMC GUI  can be installed and run
independently of each other. These programs are located in 
the ``bin'' directory.

%%%%%%%%%%%%%%%%%%%%%%%%%%%%%%
\subsection{RunMC GUI}
\label{RunMCGUI}
%%%%%%%%%%%%%%%%%%%%%%%%%%%%%%

RunMC GUI allows an interaction between the user and
RunMC MC programs. 
At present, two types of RunMC GUI are available:
a user interface based on C++ (can be executed with the
command ``runmc'') and that based on Java (the command ``jrunmc'').  
After the RunMC installation, the Java archive file ``JRunMC.jar'', which
is called by the ``jrunmc'' bash script,  
is located in the ``bin'' directory. Below we will describe 
the C++ RunMC GUI only since the Java-based GUI is very similar.

The task of RunMC GUI is to generate the output file
``project.mc'', where ``project'' is a user-defined 
name of the current calculation.  
This file contains the most 
important information for the physics analysis: 

\begin{itemize}
\item 
the type of MC model, the number of events to generate, 
the type of the initial particles, their energy 
and the structure functions.  
It also  specifies  the 
type of the output final state 
(partons, stable hadrons, charged hadrons, hadronic jets);

\item
the type of the output file: ROOT histograms or ROOT event trees. 
Two types of the ROOT trees
can be filled: the complete HEPEVT event record or only the 
final states selected via RunMC GUI 
(i.e. partons, stable hadrons etc.). 
The latter event record, which is called the RunMC ROOT tree, 
contains the  most important variables which are sufficient
to do most physics studies (jet reconstruction, event shapes, 
inclusive particle spectra etc.). 
The size of the RunMC event record
is typically four times smaller than the HEPEVT record; 

\item
the kinematic cuts applied for events or final particles/jets. 
The cuts can be set on the transverse energy 
and/or  pseudorapidity
of selected particles (or jets). 
In case of deep inelastic
scattering, the range in $Q^2$, $x$, $y$ and $W$ kinematic 
variables can be set;

\item
the presentation styles of histograms and the size of the ROOT canvas.
The histograms can be filled without normalisations or
they can be normalised to the total number of generated events.
Also, differential cross sections can be calculated 
taking into account the predicted luminosity and the histogram bin width
(this option is supported for 
one-dimensional histograms only);

\item
the pipe files used to communicate with the RunMC MC programs;

\item
the histogram definitions (the titles, the number of bins, 
the minimum and the maximum values, 
histogram dimensions).

\end{itemize} 

RunMC GUI adopts the following strategy to define the histograms:  
There are two GUI windows, ``Variables'' and ``Histograms''. 
The first window contains the names of the 
variables (with some additional comments)  
defined for a given physics project. 
The user should select
the appropriate variable and copy it to the ``Histograms'' 
window by clicking on the corresponding 
title or icon.

The variable names are divided into the three categories: 
event-based variables (characterising the event as whole), 
single-particle densities (filled for each particle/jet; 
the variable name starts  
with ``@'')  and two-particle densities 
(filled for each particle/jet pair; 
the name starts with  ``@@''). 
Such naming convention is necessary to avoid 
unnecessary loops over particles
or particle pairs if single or two-particle 
densities are not required by the user.

The histograms can also be filled in the user-defined subroutine 
``user-run.cpp''; in this case the naming  
convention for the variables discussed above is unnecessary.

Two-dimensional histograms can also be filled. 
If two one-dimensional histograms are defined, a 
two-dimensional histogram can be build from these 
two histograms using RunMC GUI.

To start the MC run, the command ``runmc'' should be executed  
from the project directory
``proj'' (or ``jrunmc'' if the user prefers the Java GUI). 
RunMC GUI allows the selection of the MC models, 
the initial particles,
their energies, the number of events, the output required, 
the project name. The histograms can be defined   
by clicking on the ``Variable'' list and 
using a  spreadsheet-like window ``Histograms''. 
The MC run can be started by executing the command "Start". 
If specific MC settings are required, rather than
the default parameters, one can create a steering file 
(with the extension ``.cards'') to change   
initial MC parameters. This can be done using 
a spreadsheet-like editor of RunMC GUI.
Next, the program performs some basic checks,
and in case of problems, 
reports them. If no
errors are found, RunMC GUI performs the following actions:

\begin{itemize}

\item 
creates a file called ``project.mc'', 
where ``project'' is the name given by the user. This
file contains the settings described above. It is always 
linked to ``.analmc.ln'' which will be 
read by the RunMC MC program;

\item 
creates a file called ``MCname.cards'', where ``MCname'' is 
the name of the current MC model 
(``pythia6'', ``herwig'', ``ariadne'', ``lepto'', 
``aroma'' ``cascade''). This steering file  
contains initial parameters of the MC models. Note that   
such file may not 
be needed if the default MC parameters are required;

\item
creates a pipe file called ``XX.stop'', where ``XX'' is a 
unique number assigned by RunMC for the current run. This file, located in the
directory ``pipes'',  contains zero if the MC run is allowed 
or ``1'' if the current run should be  terminated;

\item
creates a pipe file called ``XX.event''. This file, 
filled by the RunMC MC programs 
during the MC run, contains the number of generated MC events.
This number will be read by RunMC GUI to display the current status.

\item
executes one RunMC MC program corresponding to the selected MC model. 
The executable file has the name ``analmc.MCname'', 
where ``MCname'' is the name  of MC model. 
 
\end{itemize} 

During the event generation, the ROOT canvas can display 
the output histograms (up to eight in total), 
as well as how many events have been generated.
The output from RunMC MC 
is written to ``.analmc.log''
(a symbolic link to the ``project.log'' file).  
One can view the log file using
the ``log'' option of RunMC GUI.
At the end of the run, an additional window
can  appear with the run information
(the name of MC model, the number of the MC events processed,
the luminosity used etc.).
Possible errors are redirected to the file ``project.err'',
which is constantly monitored by RunMC GUI.

The ROOT histograms
are automatically modified at the end of the fill 
if they are required 
to be normalised to the total number of events 
or converted to differential cross sections. 
Note that  there is no need to wait until the end of the 
current run: once  
the histogram statistics is sufficient, 
one can terminate the run by clicking ``Stop'' on the GUI window.    
Histograms should be saved in the ROOT file ``project.root'' for 
further studies. 
The style of the histograms can further be modified using the ROOT
canvas editor.

All user manuals for the MC models and the RunMC user manual can be
accessed via RunMC GUI using the ``Help'' option.

%%%%%%%%%%%%%%%%%%%%%%%%%%%%%%%%%%%%%%%%%%%%
\subsection{The RunMC MC programs}
\label{RunMCMC}
%%%%%%%%%%%%%%%%%%%%%%%%%%%%%%%%%%%%%%%%%%%%

The RunMC MC programs, which are based on the FORTRAN Monte Carlo models and ROOT, 
have the genetic name  
``analmc.MCname'', where ``MCname'' is the name of the corresponding MC model.
The main function in the file ``analmc.cpp'', which is
located in the directory ``main/src'',   calls the FORTRAN subroutine
``runmcarlo'' to be described below.
This main program is based on C++ and ROOT.
The C++ code accesses the HEPEVT common block of a given MC program via a C-like 
structure. The RunMC MC program receives the initial parameters set by RunMC GUI
via the symbolic link ``.analmc.ln'' pointing to  
the file ``project.mc''. 

Each MC model has its own FORTRAN
subroutine ``runmcarlo'', which provides an interface to the FORTRAN code
of the given MC program. 
This interface program (in the file 
``RUNMC-MCname.f'')  is  located in the directory ``main/mcarlo/MCname''.
The task of the subroutine ``runmcarlo'' is to fill the HEPEVT common block.
In addition, some  
initial settings are done by accessing a C/C++ structure with the initial parameters 
defined in the ``project.mc'' file. The main function in ``analmc.cpp'' 
calls this interface subroutine and fills   
the C/C++  structure which represents the complete HEPEVT event record. 
Next, this structure was used to select the final state generated by the MC models. 
The output is copied to the  class ``HEPLIST'' which can be accessed by 
external calculations.
The class HEPLIST consists of several vectors based on the  
LorentzVector vector class (from the CLHEP library) which represents
four-momentum of a particle or a jet.
The definition of the HEPLIST class, as well as other include files, can be found
in the ``main/inc'' directory.

In addition to the FORTRAN subroutine ``RUNMC-MCname.f'', 
which communicates with the MC programs, the directory
``main/mcarlo/MCname'' contains an  additional 
FORTRAN subroutine in ``RUNMC-MCname-steer.f''. 
It was designed to read the initialisation file (``MCname.cards'') 
located in the user directory ``proj''.

The main function tries to determine which loop is necessary to use  
to fill the histograms. This decision is based on the naming convention 
for the RunMC variables described in Section~\ref{RunMCGUI}.
Note that the RunMC MC programs do  not require recompilations, since the histograms
are set via the input file. 
For a larger flexibility, the histograms 
can also be filled manually in the user-defined
file ``user-run.cpp'' (see Section~\ref{UCal}).

A several physics packages are available 
inside RunMC MC to transform the original
four-momentum vector of particles/jets to the required observable:

\begin{itemize}

\item  
the transformations provided by the physics 
vector class ``LorentzVector'' from CLHEP 
can be used, since a particle or a jet is 
represented as a general four-vector based on this class;

\item 
the event-shape calculations are available using the package developed by
M.~Iwasaki \cite{Iwasaki};

\item 
the longitudinally-invariant $k_T$ algorithm as implemented in 
C++ \cite{Butterworth:2002xg} can be used for the
jet reconstruction.
In addition to this package, the JADE and Durham jet 
algorithms are implemented according to M.~Iwasaki \cite{Iwasaki};

\item 
the Breit frame was implemented for $ep$ deep inelastic scattering.

\end{itemize}
The physics packages with the corresponding documentation are located in
the directory ``main/physics''.

The transformations based on the LorentzVector class belong to a high-level
layer of the RunMC event record. The user still can access more elementary
event records which can be used to transform them to other event classes and
physics calculations\footnote{This 
can be done in the user directory ``proj'', and packed 
in a physics-project module, see Sections~\ref{UCal} and \ref{PP}.}.   
For example, the FORTRAN HEPEVT common block can directly be accessed using
a C++ structure inside the RunMC package.   
   
%%%%%%%%%%%%%%%%%%%%%%%%%%%%%%%%%%%%%%%%%%%%
\section{User calculations}
\label{UCal}
%%%%%%%%%%%%%%%%%%%%%%%%%%%%%%%%%%%%%%%%%%%%

For a new physics calculation, the 
directory ``proj'' should be modified. 
This user directory can contain
external calculations, steering cards 
for MC initialisations, as well as  
the standard RunMC functions which are necessary to initialise and fill the histograms.

The user directory should always contain the file  ``project.mc'' created by RunMC GUI.
This file can be edited manually without the RunMC GUI program using any text editor.
One can load this file to RunMC GUI by executing the command ``runmc project.mc''
from the bash shell (or using the option ``Projects$\to$read MC'' of  RunMC GUI). 
This updates  the RunMC GUI program according to new project settings.

The directory ``proj'' can contain steering files ``MCname.cards'' 
to redefine
initial MC parameters. Such files can be created
via RunMC GUI (``MC settings'' option). For more flexibility,
the MC initialisation parameters  
can also be overwritten
by FORTRAN-coded subroutines located in the 
directory ``proj/ini''. If this is not done, the default  MC parameters
will be selected according to the RunMC option.

The histograms can  be defined in the two steps:  
First, a necessary variable should be calculated in 
the file ``user\_afill.cpp''. 
The output of this function is a pointer.  The  
output variable name should always be associated with this pointer. 
As was explained above, three types of the variables can be defined 
(for each event, for each particle/jet, for each particle/jet pairs). 
Next, the variable names 
should be specified in the file ``user-name.txt''. 
It includes the variable names to the 
list ``Variables'' accessed by  RunMC GUI.
Finally, to compile the source codes in the directory ``proj''
and to rebuild all RunMC MC programs to take into account
changes made in the project source files, 
one should type  ``make'' in the ``proj'' directory. 
All MC programs will be recompiled
and RunMC GUI will be updated with new histograms. 
Then, the ``runmc'' command  should be
executed in the user directory to start RunMC GUI.
The main advantage of this approach is that once a 
necessary variable is defined, 
new histogram definitions 
do not require the recompilation of all MC models. 

However, the approach discussed above has some limitations 
since more complicated observables cannot be filled (for example, 
if they are constructed from three or more 
particle densities). 
Therefore, to avoid this restriction, the 
RunMC histograms can also be filled using the
conventional method, i.e. in 
the function located in ``user-run.cpp''.
In this case, the initialisation of histograms is not required, as long as
the file ``project.mc'' defines which histograms should be filled and what presentation 
style should be used to fill the histograms. 

To have even more flexibility, the histograms can be initialised in the
file ``user-init.cpp'' according to the standard ROOT procedure.
In this case, each redefinition of the histograms
(the bin size, the title etc.)
requires recompilation
of all RunMC MC programs, i.e. it is necessary to run ``make'' from
the ``proj'' directory.

All RunMC MC histograms and the ROOT 
canvas can be accessed by the user using the C/C++ 
``extern''  statement.  

%%%%%%%%%%%%%%%%%%%%%%%%%%%%%%%%%%%%%%%%%%%%
\section{Physics projects}
\label{PP} 
%%%%%%%%%%%%%%%%%%%%%%%%%%%%%%%%%%%%%%%%%%%%

In order to share complicated analysis calculations 
and store them for further use in a future, 
the directory ``proj'' can be packed into an external file.
Such a file, which contains  
the zipped directory ``proj'',  is called the RMC project file. 
It has the extention ``rmc''. For example, ``project.rmc'' is 
the  RMC file which has a user-defined name ``project''.
The ``proj'' directory inside of this RMC file has one and
only one file ``project.mc'' with RunMC GUI settings. 
As was discussed in Section~\ref{UCal}, this directory may contain
some user-defined external functions, libraries, make files 
and MC steering files.

RunMC GUI can read such project RMC file 
automatically using the option. 
``Projects$\to$load RMC''. The RunMC GUI program does this 
in a few steps: 

\begin{itemize}
\item  
unzips this file and 
replaces the content of the directory ``proj'' with new files from
``project.rmc''. The old user directory 
is zipped and copied to the  directory ``main/tmp'';

\item
recompiles the project (i.e. it executes ``make'' 
from the directory ``proj''). 
RunMC GUI displays the status bar indicating 
the compilation process;

\item
updates the GUI window with new variables, histograms and options. 

\end{itemize}  

After a short time (typically 30-60 sec depending on the CPU), 
the new project is loaded to RunMC GUI.
To run it, the user should select ``Start$\to$Run'', and after
making sure that the project can successfully be executed, make necessary modifications
via RunMC GUI or in the source files.

To save the project,
one should select ``Projects$\to$archive RMC''. This   creates 
the file ``project.rmc'' from the content of the directory ``proj''. 
Then this file will be copied to the directory ``archive''.

At present, several RunMC project files 
are available on the Web \cite{S.Chekanov} 
(they are also included in  
the directory ``archive'' of RunMC):

\begin{itemize}
\item  
the default project. Only pre-installed variables can be included 
in the calculations.
All user-defined functions in the directory ``proj'' are dummy; 

\item 
HERA kinematic variables calculated in user-defined functions ($Q^2$, $x$ etc.);

\item  
jets at HERA using the longitudinally-invariant 
$k_T$ algorithm in the Breit frame;

\item  
jets at LHC.
As for the previous module, the jet cross sections are reconstructed using  
the longitudinally-invariant $k_T$ algorithm; 

\item
jet cross sections at LHC for parton and hadron levels.
This project contains the calculation of hadronisation
corrections by taking the ratio of these cross
sections. It also illustrates how to initialise and to fill
histograms in the function ``user-run.cpp'';

\item
calculations of the  $D^*$ cross sections in $ep$ collisions at HERA;

\item
calculations of the cross sections for strange-particle production in $ep$
collisions at HERA; 

\item
the HZTOOL package~\cite{hztool};   

\item  
the event-shape variables in $e^+e^-$ at NLC energies; 

\item
several examples of how to visualise tracks and the $k_T$ jets
in 3D for a single MC event ($e^+e^-$, $ep$, $pp$ collisions). 
They use a simple imitation of
the magnetic field using the  example taken from the
ROOT project \cite{Brun:1997pa,Brun:2003ga}.
 
\end{itemize}

The RMC project files discussed above only
illustrate how to set up and to  
develop new physics calculations in the RunMC framework.
For practical applications, these examples should be modified.

%%%%%%%%%%%%%%%%%%%%%%%%%%%%%%%%%%%%%%%%%%%%
\section{Running in the background}
%%%%%%%%%%%%%%%%%%%%%%%%%%%%%%%%%%%%%%%%%%%%

Due to a complete independence of RunMC GUI and RunMC MC 
programs, one can run jobs 
in the background without GUI or any pop-up window. 
This can be done using the following steps:

\begin{itemize}
\item
edit the source files in the directory ``proj'', 
or unzip a RMC project file to this directory. 
Recompile the project by typing ``make'' from the directory ``proj'';  

\item
make sure that the file ``project.mc'' is linked to ``.analmc.ln''.
If you do not want ROOT pop-up canvas with histograms, 
a corresponding option should be set
in the file ``project.mc''
(see the line ``OPTIONS'' in ``project.mc'', you may use 
RunMC GUI first to find out which  RunMC parameter should be  
modified); 

\item 
execute the file ``analmc.MCname'' corresponding to 
the particular MC model. 
Normally, the program should ask first
to create a pipe file with ``0'' in the directory ``pipes''
which allows this run.
\end{itemize}

The MC run can be terminated
at any time by setting ``1'' in the pipe file.
All histograms should be saved in the file ``project.root''.

%%%%%%%%%%%%%%%%%%%%%%%%%%%%%%%%%%%%%%%%%%%%
\section{RunMC ROOT tree analyser}
%%%%%%%%%%%%%%%%%%%%%%%%%%%%%%%%%%%%%%%%%%%%

In addition to the standard functionality of 
the MC event simulation, 
RunMC GUI can also use ROOT trees as the input for physics calculations. 

The ROOT tree can be generated by selecting  
an appropriate option via RunMC GUI, either ``HEPEVT'' or ``RUNMC'', 
in addition or instead of the ROOT histogram option. 
Then, the MC events should be generated as usual, 
but this time a ROOT tree with the 
extention ``.rtup'' or ``.htup'' will be created. Then, 
RunMC can run over this ROOT tree  
if, instead of the MC model, 
the option ``RUNMC'' or `HEPEVT'' is selected.
Several ROOT trees  can automatically be  
included in the analysis
as long as they are of
the same type and located in the same directory.

The analysis of the ROOT trees is very similar to 
the standard run over MC events.
The external 
RMC files can be used to define new variables and to specify 
the output histograms. 
The only difference is that
the cross section calculations are not possible at present.

The main advantage of the ROOT tree analyser is 
that physics calculations and histograms can be produced  
significantly faster than when MC models are used 
to generate and to fill histograms at the same time. 
In case of the ROOT tree, 
RunMC can fill histograms by a factor $\sim$10--15 faster,
thus the RMC project files can be
validated and analysed more efficiently.

With this additional feature, 
RunMC can also be used to analyse experimental data if 
the event record is converted to  
the appropriate ROOT tree.  As for the 
Monte Carlo models, the data analysis can be performed
using the RMC project files.

%%%%%%%%%%%%%%%%%%%%%%%%%%%%%%%%%%%%%
\section*{Acknowledgements}
%%%%%%%%%%%%%%%%%%%%%%%%%%%%%%%%%%%%%
Argonne National Laboratory's work was supported under
U.S. Department of Energy contract W-31-109-Eng-38.

%%%%%%%%%%%%%%%%%%%%%% references %%%%%%%%%%%%%%%%%%%%%%%%%%%%%%
\bibliography{biblio}

\begin{thebibliography}{10}
\expandafter\ifx\csname url\endcsname\relax
  \def\url#1{\texttt{#1}}\fi
\expandafter\ifx\csname urlprefix\endcsname\relax\def\urlprefix{URL }\fi

\bibitem{Bertini:2000uh}
M.~Bertini, L.~L\"onnblad, T.~Sj\"{o}strand, {\sc PYTHIA} version 7: A
  proof-of-concept version, Comput. Phys. Commun. 134 (2001) 365.

\bibitem{ThePEG}
L.~L\"onnblad, {\sc ThePEG}: Toolkit for high energy physics event generation.
\newline\urlprefix\url{http://www.thep.lu.se/ThePEG/}

\bibitem{paw}
Application-Software-Group, {\sc PAW:} Physics Analysis Workstation.
\newline\urlprefix\url{http://wwwasd.web.cern.ch/wwwasd/paw/}

\bibitem{hbook}
R.~Brun, D.~Lienart, {\sc HBOOK:} statistical analysis and histogramming,
  CERN-Y250.

\bibitem{Lonnblad:1994kt}
L.~L\"onnblad, {\sc CLHEP}: A project for designing a {\sc c++} class library
  for high-energy physics, Comput. Phys. Commun. 84 (1994) 307.

\bibitem{Fischler:2000em}
M.~Fischler, A.~Pfeiffer, {\sc CLHEP} - new developments and directions, in:
  The proceedings of International Conference on Computing in High Energy
  Physics and Nuclear Physics (CHEP 2000), Padova, Italy, 7-11 Feb 2000.
\newline\urlprefix\url{http://wwwasd.web.cern.ch/wwwasd/lhc++/clhep/}

\bibitem{Brun:1997pa}
R.~Brun, F.~Rademakers, {\sc ROOT}: An object oriented data analysis framework,
  Nucl. Instrum. Meth. A389 (1997) 81.
\newline\urlprefix\url{http://root.cern.ch/}

\bibitem{Brun:2003ga}
R.~Brun, F.~Rademakers, P.~Canal, M.~Goto, Root status and future developments,
  ECONF C0303241 (2003) MOJT001.

\bibitem{JetWeb}
J.~M. Butterworth, S.~Butterworth, {\sc JetWeb:} a {\sc www} interface and
  database for monte carlo tuning and validation, Comput. Phys. Commun. 153
  (2003) 164.

\bibitem{Hirabayashi}
T.~Hirabayashi, The wide studio project.
\newline\urlprefix\url{http://www.widestudio.org}

\bibitem{Sjostrand:2001yu}
T.~Sj\"{o}strand, L.~L\"onnblad, S.~Mrenna, Pythia 6.2: Physics and manual.

\bibitem{herwig}
G.~Corcella, et~al., {\sc HERWIG 6: } an event generator for hadron emission
  reactions with interfering gluons including supersymmetric processes, JHEP
  0101 (2001) 10.

\bibitem{Lonnblad:1992tz}
L.~L\"onnblad, {\sc ARIADNE} version 4: A program for simulation of qcd
  cascades implementing the color dipole model, Comput. Phys. Commun. 71 (1992)
  15.

\bibitem{Ingelman:1996mq}
G.~Ingelman, A.~Edin, J.~Rathsman, {\sc LEPTO 6.5} - a monte carlo generator
  for deep inelastic lepton-nucleon scattering, Comput. Phys. Commun. 101
  (1997) 108.

\bibitem{Ingelman:1996mv}
G.~Ingelman, J.~Rathsman, G.~Schuler, {\sc AROMA 2.2} - a monte carlo generator
  for heavy flavour events in $ep$ collisions, Comput. Phys. Commun. 101 (1997)
  135.

\bibitem{Jung:2001hx}
H.~Jung, The {\sc ccfm} monte carlo generator {\sc cascade}, Comput. Phys.
  Commun. 143 (2002) 100.

\bibitem{phojet}
R.~Engel, {\sc PHOJET}.
\newline\urlprefix\url{http://www-ik.fzk.de/~engel/phojet.html}

\bibitem{rapgap}
H.~Jung, Hard diffractive scattering in high energy ep collisions and the monte
  carlo generator {\sc rapgap}, Comm. Phys. Commun. 86 (1995) 147.

\bibitem{genetic}
C.~Bal\'{a}zs, et~al., The {\sc qcd/sm} working group. {\sc monte carlo}, in:
  P.~Aurenche, et~al. (Eds.), Proceedings of the Workshop on Physics at TeV
  Colliders, Les Houches, France, 2001, p.~95.
\newline\urlprefix\url{http://wwwlapp.in2p3.fr/conferences/LesHouches/Houches2%
001/}

\bibitem{Iwasaki}
M.~Iwasaki, {\sc Jet-EventShape-Finders} package.
\newline\urlprefix\url{ftp://ftp.slac.stanford.edu/groups/lcd/Physics_tools/}

\bibitem{Butterworth:2002xg}
J.~M. Butterworth, J.~P. Couchman, B.~E. Cox, B.~M. Waugh, {\sc KtJet}: A
  c++implementation of the $k_t$ clustering algorithm, Comput. Phys. Commun.
  153 (2003) 85.

\bibitem{S.Chekanov}
S.~Chekanov, {\sc RunMC} - {\sc c++} object-oriented framework for running
  monte carlo models for linux and windows.
\newline\urlprefix\url{http://www.desy.de/~chekanov/runmc}

\bibitem{hztool}
J.~Bromley, et~al., {\sc HZTOOL:} a package for {\sc monte carlo} - data
  comparisons at hera, in: W.~Buchmuller, G.~Ingelman (Eds.), Future Physics
  and HERA, Vol.~2, Hamburg, Germany, 1996-1997, p. 611.
\newline\urlprefix\url{http://hztool.hep.ucl.ac.uk/}

\end{thebibliography}
\end{document}